\documentstyle[12pt]{article}
\begin{document}
\setcounter{page}{1}
\thispagestyle{empty}

\centerline{\normalsize hep-ph/9410409 \hfill OITS-554}
\centerline{\normalsize \hfill September 1994}

\vspace{15mm}

\begin{center} {\bf QUARK MASS CORRECTIONS TO THE Z BOSON DECAY RATES}
\footnote{The part of this work has been done in collaboration with
                                                         D.~E.~Soper.}

\vspace{1cm}
               {\bf Levan R. Surguladze} \\
\vspace{2mm}
{\it Institute of Theoretical Science, University of Oregon\\
                  Eugene, OR 97403, USA}
\end{center}

\vspace{3mm}

\begin{abstract}
The results of perturbative QCD evaluation of the $\sim m_f^2/M_Z^2$
contributions to $\Gamma_{Z\rightarrow b\overline{b}}$ and
$\Gamma_{Z\rightarrow hadrons}$ for the quark masses $m_f\ll M_Z$
are presented. The recent results due to the combination of
renormalization group constraints and the results of several other
calculations are independently confirmed by the direct computation.
Some existing confusion in the literature is clarified. In addition,
the calculated $O(\alpha_s^2)$ correction to the correlation function
in the axial channel is a necessary ingredient for the yet uncalculated
axial part of the $O(\alpha_s^3)$ mass correction to the Z decay rates.
The results can be applied to the $\tau$ hadronic width.
\end{abstract}

\vspace{1cm}

   Recent analyses show that the LEP result for the branching fraction
$\Gamma_{Z\rightarrow b\overline{b}}/\Gamma_{Z \rightarrow hadrons}
= 0.2208\pm 0.0024$ \cite{LEP}
differs (is larger than) the Standard Model prediction \cite{Knirev}
by 2-2.5 $\sigma$ with the top mass at around 174 GeV \cite{CDF}.
Although the search for implications of this fact beyond the Standard
Model is already began (see, e.g., \cite{BSM}), a further analysis
within the Standard Model is still an important issue.

   Briefly, a current state of the perturbative QCD evaluation
of the $\Gamma_{Z \rightarrow hadrons}$ and related quantities is as
follows. The QCD contributions are evaluated to $O(\alpha_s^3)$ in
the limits $m_f=0$ ($f=u,d,s,c,b$) and $m_t\rightarrow \infty$ \cite{Rs}.
The leading correction to the above results due to the $O(\alpha_s^3)$
triangle anomaly type diagrams with the virtual top quark has also been
calculated in the limit $m_t\rightarrow \infty$ \cite{Lar}.
The calculations to $O(\alpha_s^2)$ are more complete.
Indeed, the electron positron annihilation R-ratio was
calculated long ago \cite{Dine} in the limits $m_f=0$ ($f=u,d,s,c,b$) and
$m_t\rightarrow \infty$. The corrections due to the large top-bottom
mass splitting have been evaluated in \cite{Kni,Chettri} and thus
the axial channel, applicable to $\Gamma_{Z \rightarrow hadrons}$,
has also been validated. The $O(\alpha_s^2)$ effects of the virtual heavy
quark in the decays of the Z-boson have been evaluated in \cite{Sop,Knivert}.
The same effect has been studied previously in \cite{Chetvirt} in the
limit $m_t \rightarrow \infty$.
The present knowledge of the high order QCD and electroweak corrections
to $\Gamma_{Z \rightarrow hadrons}$ are summarized in the review article
\cite{Knirev}, providing all essential details.

    The subject of the present work is the correction due to the
nonvanishing ``light'' quark masses. The following discussion will
be for the quark of flavor $f$ ($f\neq t$) in general. However, in fact,
only $f=b$ is a relevant case and the masses of $u,d,s$ and $c$ quarks
can safely be ignored at the Z mass scale. Note that the corrections
$\sim \alpha_s^2m_b^2/M_Z^2$ for the vector and axial parts
of the $Z\rightarrow b\overline{b}$ decay rate were obtained
in \cite{Chetm2V} and \cite{Chetm2A} correspondingly.
Those evaluations are based on an indirect approach, using
the renormalization group constraints and the results of earlier
calculations of the correlation functions in the vector \cite{GKL,Su}
and scalar channels \cite{GKLS}. One of the aims of the present work
is to obtain the quark mass corrections to
$\Gamma_{Z\rightarrow q_f\overline{q}_f}$
in both channels by a direct calculation and to check the method
and the results used in \cite{Chetm2V,Chetm2A} at once.
Moreover, it should be stressed that there is a disagreement between
the results of \cite{GKL} and \cite{Su}.
\footnote{Numerically the disagreement is not large.
In \cite{Katcom} the result of \cite{Su} was confirmed.
However, as it will be shown in the present paper, the initial result
of \cite {GKL} turns out to be correct. Also note that unfortunately,
in \cite{Chetm2V}, the result of \cite{Su} was used.}

The quantity $R_Z$ is defined as the ratio of the hadronic and electronic
Z widths
\begin{equation}
R_{Z}=\frac
    {\sum_{f=u,d,s,c,b}\Gamma_{Z\rightarrow q_{f}\overline{q}_f}}
    {\Gamma_{Z\rightarrow e^{+}e^{-}}}.
\label{Rz}
\end{equation}
The partial decay width can be evaluated as the imaginary part
\begin{equation}
\Gamma_{Z\rightarrow q_{f}\overline{q}_f}
   =-\frac{1}{M_Z}\mbox{Im}\Pi(m_f,s+i0)\biggr|_{s=M_Z^2},
 \label{ImPi}
\end{equation}
where the function $\Pi$ is defined through a correlation function
of two flavor diagonal quark currents
\begin{equation}
 i\int d^4x e^{iqx}<Tj^f_{\mu}(x)j^f_{\nu}(0)>_{0}
   =g_{\mu\nu}\Pi(m_f,Q^2)-Q_{\mu}Q_{\nu}\Pi'(m_f,Q^2).
\label{PI}
\end{equation}
Here, $Q^2$ is a large ($\sim -M_Z^2$) Euclidean momentum.
According to the Standard Model, the neutral weak current of quark
coupled to Z boson is
\begin{equation}
j_{\mu}^{f}=\biggl(\frac{G_FM_Z^2}{2\sqrt{2}}\biggr)^{1/2}
      (g^{V}_{f}\overline{q}_f\gamma_{\mu}q_f
      +g^{A}_{f}\overline{q}_f\gamma_{\mu}\gamma_5q_f),
\label{j}
\end{equation}
where the electroweak vector and axial couplings are defined in a standard
way:
\begin{displaymath}
g_f^V=2I^{(3)}_f-4e_f\sin^2\Theta_W, \hspace{5mm} g_f^A=2I^{(3)}_f.
\end{displaymath}
The $\Pi$-function may be decomposed into vector and axial parts
\begin{equation}
\Pi(m_f,Q^2)=\Pi^V(m_f,Q^2)+\Pi^A(m_f,Q^2).
\label{Pisum}
\end{equation}

    The Feynman diagrams that contribute in $\Pi$
to $O(\alpha_s^2)$ are shown in Fig.~1.

\vspace{5cm}

\begin{center}
{\bf Fig.1} Feynman diagrams contributing in the correlation function
to $O(\alpha_s^2)$. The cut diagrams contribute in
$\Gamma_{Z\rightarrow q_{f}\overline{q}_f}$ at the same order.
\end{center}

\vspace{3mm}

The effects of the last two diagrams in Fig.~1 (so called triangle
anomaly type diagrams)
have been studied in \cite{Kni,Chettri} and will not be considered
here. In the first diagram, the shaded bulb includes any interactions
of quarks and gluons (or ghosts) allowed in QCD and the dots cover any
number of gluon propagators that gives one-, two- and three-loop topologies.
The crosses denote current vertices corresponding to the vector or the axial
parts of the current (\ref{j}).

Because the problem scale ($\sim M_Z$) is much larger than the
quark masses, the following expansion is legitimate:
\begin{equation}
\Pi^{V/A}(m_f,m_v,Q^2)
   =\Pi_1^{V/A}(Q^2)+\frac{m_f^2}{Q^2}\Pi_{m_f^2}^{V/A}(Q^2)
      +\sum_{v=u,d,s,c,b}\frac{m_{v}^2}{Q^2}\Pi_{m_v^2}^{V/A}(Q^2)
        +\cdots
\label{Piexpan}
\end{equation}
The last term in the above expansion is due to the certain topological
types of three-loop diagrams containing virtual fermionic loop.
The effects of the virtual top quark in the decays
of the Z-boson have been studied in \cite{Sop,Knivert,Chetvirt} and
will not be discussed here. The period in eq.(\ref{Piexpan})
covers the terms $\sim m_f^4/Q^4$ and higher orders. Those terms
at $-Q^2=M_Z^2$ for the Z decay are heavily suppressed and can safely be
ignored.

   The expansion coefficients $\Pi_i$ can be calculated in a way similar
to the one used in \cite{H} for the evaluation of the fermionic decay
rates of the Higgs boson. In fact, the whole calculational
procedure can be combined in one equation (in the limit
$m_t\rightarrow \infty$)
\newpage
\begin{eqnarray}
\lefteqn{\hspace{-7mm}\Gamma_{Z\rightarrow q_{f}\overline{q}_f}
 =-\sum_{i=V,A}\sum_{\stackrel{n,k=0,1}{n+k\leq 1}}
  \frac{1}{(2n)!(2k)!}} \nonumber\\
  && \quad \hspace{-7mm}
 \frac{1}{M_Z}
 \mbox{Im}\biggl\{Z_m^{2(n+k)}
  m_f^{2n}m_v^{2k}
  \biggl[\biggl(\frac{d}{dm_f^B}\biggr)^{2n}
         \biggr(\frac{d}{dm_v^B}\biggr)^{2k}
     \Pi^i(\alpha_s^B,m_f^B,m_v^B,s+i0)\biggr]
     _{\stackrel{m_f^B=m_v^B=0}
                {\alpha_s^B\rightarrow Z_{\alpha}\alpha_s}}
                     \biggr\}_{s=M_Z^2}
\label{main}
\end{eqnarray}
where B labels the unrenormalized quantities.
$Z_m$ and $Z_{\alpha}$ are the $\overline{MS}$
renormalization constants of the quark mass and the strong coupling
correspondingly and can be found, for instance,
in \cite{H}. The summation over the virtual quark
flavors $v=u,d,s,c,b$ is assumed in $\Pi^i$.
Note that the introduction of the so called
$D$-function (see, e.g., \cite{Chetm2V,Chetm2A}), which is the
$\Pi$-function  differentiated with respect of $Q^2$, is not necessary
in this calculation. Note also that eq.(\ref{main}) does not include
important effects from the last two diagrams in Fig.~1
(evaluated in \cite{Kni,Chettri}) and the virtual top quark effects
(evaluated in \cite{Sop,Knivert,Chetvirt}).

In the $\overline{MS}$ \cite{MSB} analytical calculations of the one-,
two-, and three-loop dimensionally regularized \cite{drg} Feynman
diagrams, the FORM \cite{FORM} program HEPLoops \cite{HEPL} is used.

For the massless limit coefficients $\Pi_1^{V/A}$ in the expansion
(\ref{Piexpan}), the known results are obtained.
The perturbative expansion of the $\sim m_f^2$ part in the r.h.s.
of eq.(\ref{Piexpan}) has the form
\begin{eqnarray}
\lefteqn{\hspace{-8mm}\frac{m_f^2(\mu)}{Q^2}
           \Pi^{V/A}_{m_f^2}(\alpha_s(\mu),Q^2)
       + \sum_{v=u,d,s,c,b}\frac{m_v^2(\mu)}{Q^2}
       \Pi^{V/A}_{m_v^2}(\alpha_s(\mu),Q^2) }\nonumber\\
 && \quad \hspace{10mm}
 =\frac{G_FM_Z^2}{8\sqrt{2}\pi^2}g_f^{V/A}
 \sum_{i=0}^{3}\sum_{j=0}^{i+1}
 \biggl(\frac{\alpha_s(\mu)}{\pi}\biggr)^i
  \log^j\frac{\mu^2}{Q^2}
 \biggl(m_f^2(\mu)d_{ij}^{V/A}+e_{ij}\sum_{v=u,d,s,c,b}m_v^2(\mu)\biggr).
\label{PIpexpf}
\end{eqnarray}
The coefficients $e_{ij}$ are the same in both channels
for obvious reasons. Moreover, they get nonzero values
starting at the three-loop level ($i\geq 2$). The summation
index $j$ runs from zero to $i+1$ since, in general, the maximum
power of the pole that can be produced by a multiloop Feynman diagram
is equal to the number of loops.

    The direct computation of all relevant one-, two- and three-loop
Feynman diagrams for the standard QCD with the SU$_c(3)$ gauge group
gives in the vector channel:
\begin{displaymath}
d_{00}^V=-6;
\end{displaymath}
\begin{equation}
d_{10}^V=-16,  \hspace{5mm}  d_{11}^V=-12;
\label{dijV}
\end{equation}
\begin{displaymath}
d_{20}^V=-\frac{19691}{72}-\frac{124}{9}\zeta(3)
 +\frac{1045}{9}\zeta(5)
         +N_f\frac{95}{12},
\hspace{5mm} d_{21}^V=-\frac{253}{2}+N_f\frac{13}{3},
\hspace{5mm} d_{22}^V=-\frac{57}{2}+N_f;
\end{displaymath}
\begin{displaymath}
e_{00}=e_{1j}=0, \hspace{5mm}
e_{20}=\frac{32}{3}-8\zeta(3), \hspace{5mm} e_{21}=e_{22}=0.
\end{displaymath}
Note that $d_{i,i+1}^V=e_{i,i+1}=0$, because the highest poles
cancel at each order after the summation of Feynman graphs within each
gauge invariant set. This is the consequence of the conservation
of current. The above results fully confirm the findings of \cite{GKL}
(see also \cite{Chettau}). On the other hand, the $\zeta(3)$ coefficient
in $d_{20}$ disagrees with the incorrect one presented in \cite{Su}.
\footnote{in contrary to the previous belief (see, e.g., comments in
\cite{Chetm2V,Katcom}), that the result of \cite{GKL} has been corrected
in \cite{Su}.}
Fortunately, the numerical difference is small. The error in \cite{Su}
is due to the misprint in the $\zeta(3)$ term in the result for the
three-loop nonplanar type diagram.

It can be shown that the vector part of the l.h.s. of
eq.(\ref{PIpexpf}) is invariant under the renormalization group
transformations and obeys the renormalization group equation
\begin{equation}
\biggl(\mu^2\frac{\partial}{\partial\mu^2}
   +\beta(\alpha_s)\alpha_s\frac{\partial}{\partial\alpha_s}
   -\gamma_m(\alpha_s) \sum_{l=f,v}
        m_l\frac{\partial}{\partial m_l} \biggr)
    [\mbox{\footnotesize vector part of the l.h.s of eq.(\ref{PIpexpf})}]=0,
\label{RG}
\end{equation}
where $\beta(\alpha_s)$ and $\gamma_m(\alpha_s)$ are the QCD
$\beta$-function and the quark mass anomalous dimension correspondingly.
{}From eqs.(\ref{PIpexpf},\ref{RG}), it is straightforward (similar to
\cite{H})
to get the $O(\alpha_s^3)$ logarithmic coefficients in the vector channel
\begin{displaymath}
d_{31}^V=2(\beta_0+\gamma_0)d_{20}^V
  +(\beta_1+2\gamma_1)d_{10}^V+2\gamma_2d_{00}^V,
\end{displaymath}
\begin{equation}
d_{32}^V=(\beta_0+\gamma_0)[2\gamma_1d_{00}^V
           +(\beta_0+2\gamma_0)d_{10}^V]
            +(\beta_1+2\gamma_1)\gamma_0d_{00}^V,
\label{dij3}
\end{equation}
\begin{displaymath}
d_{33}^V=\frac{2}{3}\gamma_0(\beta_0+\gamma_0)
                         (\beta_0+2\gamma_0)d_{00}^V,
 \hspace{5mm} d_{34}=0,
\end{displaymath}
\begin{displaymath}
e_{31} = 2(\beta_0+\gamma_0)e_{20}, \hspace{5mm}
e_{32}=e_{33}=e_{34}=0.
\end{displaymath}
The known perturbative coefficients $\beta_n$ and $\gamma_n$
of the QCD $\beta$-function and the quark mass anomalous dimension
$\gamma_m$ with the proper normalization factors may be found, e.g., in
\cite{H}. The only missing coefficients at the
$O(\alpha_s^3)$ in eq.(\ref{PIpexpf}) for the vector channel
are the nonlogarithimc terms $d_{30}^V$ and $e_{30}$. However, these
terms have a zero imaginary part and do not contribute in the decay
rate to $O(\alpha_s^3)$.

In the axial channel, the direct computation of the relevant
one-, two- and three-loop Feynman graphs yields:
\begin{displaymath}
d_{00}^A=\frac{6}{\varepsilon}+6, \hspace{5mm}
d_{01}^A=6;
\end{displaymath}
\begin{displaymath}
d_{10}^A=-\frac{6}{\varepsilon^2}+\frac{5}{\varepsilon}
         +\frac{107}{2}-24\zeta(3),
 \hspace{5mm}  d_{11}^A=22, \hspace{5mm} d_{12}^A=6;
\end{displaymath}
\begin{eqnarray}
\lefteqn{\hspace{-22mm}
 d_{20}^A= \frac{19}{2\varepsilon^3}-\frac{99}{4\varepsilon^2}
         +\biggl(\frac{455}{36}-\zeta(3)\biggr)\frac{1}{\varepsilon}
         +\frac{3241}{6}-387\zeta(3)-\frac{3}{2}\zeta(4)
         +165\zeta(5) } \nonumber\\
 && \quad \hspace{30mm}
         -N_f\biggl(
          \frac{1}{3\varepsilon^3}-\frac{5}{6\varepsilon^2}
         +\frac{2}{3\varepsilon}
         +\frac{857}{36}-\frac{32}{3}\zeta(3) \biggr),
\label{dijA}
\end{eqnarray}
\begin{displaymath}
d_{21}^A=\frac{8221}{24}-117\zeta(3)
        -N_f\biggl(\frac{151}{12}-4\zeta(3)\biggr),
\hspace{5mm} d_{22}^A=\frac{155}{2}-\frac{8}{3}N_f,
\hspace{5mm} d_{23}^A=\frac{19}{2}-\frac{1}{3}N_f,
\end{displaymath}
where $\varepsilon=(4-D)/2$ is the deviation of the dimension of
space time from its physical value 4 within the dimensional regularization
\cite{drg}. Note that the nonlogarithmic terms in $d_{i0}^A$ contain poles,
which cannot be removed by the renormalization of the quark mass and the
coupling and have to be subtracted independently. However, one may not
worry about those poles, since the imaginary parts of nonlogarithmic
terms vanish anyway and do not contribute in the decay rate.

The above mass corrections to the three-loop correlation function
of the axial vector quark currents are the new results of the present
paper.

One may try to use the renormalization group arguments to obtain
the $O(\alpha_s^3)$ logarithmic terms similarly to the vector channel
( eq.(\ref{dij3}) ). However, to do so, the knowledge of the
$O(\alpha_s^3)$ anomalous dimension is necessary along the calculated
$d_{20}^A$ coefficient. In fact, as it was discovered in \cite{Chetm2A},
using the axial Ward identity, this anomalous dimension can be connected
to the correlation function of the quark scalar currents which, however,
is also known only to $O(\alpha_s^2)$ \cite{GKLS,H}.

{}From eqs.(\ref{Piexpan})-(\ref{dijV}), (\ref{dij3}) and
(\ref{dijA}), one obtains for the decay rate
\begin{equation}
\Gamma_{Z\rightarrow q_{f}\overline{q}_f}
 =\frac{G_FM_Z^3}{8\sqrt{2}\pi} \sum_{k=V,A}g_f^{k}
 \sum_{i=0}^{3}\sum_{j=0}^{i}
 \biggl(\frac{\alpha_s(\mu)}{\pi}\biggr)^i
  \log^j\frac{\mu^2}{M_Z^2}
 \biggl(a_{ij}^k+\frac{m_f^2(\mu)}{M_Z^2}b_{ij}^k
        +\sum_{v=u,d,s,c,b}\frac{m_v^2(\mu)}{M_Z^2}c_{ij}^k\biggr)
\label{Zmain}
\end{equation}
The massless limit coefficients $a_{ij}^{V/A}$ are calculated
up to $O(\alpha_s^3)$ \cite{Rs}. The coefficients $b_{ij}$ read:
\begin{displaymath}
b_{00}^V=0, \hspace{5mm} b_{00}^A=-6;
\end{displaymath}
\begin{displaymath}
\hspace{1mm} b_{10}^V=12, \hspace{5mm} b_{11}^V=0,
\hspace{5mm} b_{10}^A=-22,
\hspace{5mm} b_{11}^A=-12;
\end{displaymath}
\begin{equation}
b_{20}^V=\frac{253}{2}-\frac{13}{3}N_f,
\hspace{5mm} b_{21}^V=57-2N_f,  \hspace{5mm} b_{22}^V=0,
\label{bij}
\end{equation}
\begin{displaymath}
b_{20}^A=-\frac{8221}{24}+57\zeta(2)+117\zeta(3)
     +N_f\biggl(\frac{151}{12}-2\zeta(2)-4\zeta(3)\biggr),
\hspace{1mm} b_{21}^A=-155+\frac{16}{3}N_f,
\hspace{1mm} b_{22}^A=-\frac{57}{2}+N_f
\end{displaymath}
The $O(\alpha_s^3)$ coefficients for the vector part read
\begin{eqnarray}
\lefteqn{\hspace{-2cm}
b_{30}^V=2522-\frac{855}{2}\zeta(2)+\frac{310}{3}\zeta(3)
-\frac{5225}{6}\zeta(5)
-N_f\biggl(\frac{4942}{27}-34\zeta(2)+\frac{394}{27}\zeta(3)
     -\frac{1045}{27}\zeta(5)\biggr)}\nonumber\\
 && \hspace{85mm}
 +N_f^2\biggl(\frac{125}{54}-\frac{2}{3}\zeta(2)\biggr)
\label{bij3}
\end{eqnarray}
\begin{displaymath}
b_{31}=\frac{4505}{4}-\frac{175}{2}N_f+\frac{13}{9}N_f^2,
\hspace{3mm}
b_{32}=\frac{855}{4}-17N_f+\frac{1}{3}N_f^2,
\hspace{3mm}
b_{33}=0.
\end{displaymath}
The coefficients $c_{ij}$ in both channels are
\begin{equation}
c_{1j}=c_{2j}=0, \hspace{5mm}
c_{30}=-80+60\zeta(3)+N_f\biggl(\frac{32}{9}-\frac{8}{3}\zeta(3)\biggl),
\hspace{5mm} c_{31}=c_{32}=c_{33}=0.
\label{cij}
\end{equation}
The evaluation of $O(\alpha_s^3)$ coefficients $b_{3j}^A$
for the axial part requires the corresponding four-loop calculations.
The $\zeta(2)$ terms in the above coefficients are due to the
imaginary part of the term $\sim \log^3(\mu^2/s)$, which appears in the
$O(\alpha_s^3)$ coefficients of the correlation function.

Taking $\mu=M_Z$, $N_f=5$ and recalling the known massless
limit coefficients \cite{Rs}, one obtains numerically
\begin{eqnarray}
\lefteqn{\hspace{-1mm}\Gamma_{Z\rightarrow q_{f}\overline{q}_f}
 =\frac{G_FM_Z^3}{8\sqrt{2}\pi}
\biggl\{ }\nonumber\\
 && \quad \hspace{-1mm}
(2I^{(3)}_f-4e_f\sin^2\Theta_W)^2
\biggl[\biggl(1+\frac{2m_f^2(M_Z)}{M_Z^2}\biggr)
\sqrt{1-\frac{4m_f^2(M_Z)}{M_Z^2}} \nonumber\\
 && \quad \hspace{15mm}
+\frac{\alpha_s(M_Z)}{\pi}\biggl(1+12\frac{m_f^2(M_Z)}{M_Z^2}\biggr)
                                                      \nonumber\\
 && \quad \hspace{15mm}
+\biggl(\frac{\alpha_s(M_Z)}{\pi}\biggr)^2
\biggl(1.4092+104.833\frac{m_f^2(M_Z)}{M_Z^2}\biggr) \\
 && \quad \hspace{15mm}
+\biggl(\frac{\alpha_s(M_Z)}{\pi}\biggr)^3
\biggl(-12.805+547.879\frac{m_f^2(M_Z)}{M_Z^2}
       -6.12623\sum_{v=u,d,s,c,b}\frac{m_v^2(M_Z)}{M_Z^2}\biggr)\biggr]
                                                         \nonumber\\
 && \quad \hspace{-2mm}
+(2I^{(3)}_f)^2\biggl[\biggl(1-\frac{4m_f^2(M_Z)}{M_Z^2}\biggr)^{3/2}
                                                         \nonumber\\
 && \quad \hspace{15mm}
+\frac{\alpha_s(M_Z)}{\pi}\biggl(1-22\frac{m_f^2(M_Z)}{M_Z^2}\biggr)
                                                      \nonumber\\
 && \quad \hspace{15mm}
+\biggl(\frac{\alpha_s(M_Z)}{\pi}\biggr)^2
\biggl(1.4092-85.7136\frac{m_f^2(M_Z)}{M_Z^2}\biggr) \nonumber\\
 && \quad \hspace{15mm}
+\biggl(\frac{\alpha_s(M_Z)}{\pi}\biggr)^3
\biggl(-12.767+(\mbox{{\footnotesize unknown}})\frac{m_f^2(M_Z)}{M_Z^2}
       -6.12623\sum_{v=u,d,s,c,b}\frac{m_v^2(M_Z)}{M_Z^2}\biggr)\biggr]
\biggr\}, \nonumber
\label{Znum}
\end{eqnarray}
where for the Born terms, their well known exact expressions
\cite{Knirev} are used. (These terms have once again been reevaluated here.)
It should be stressed that, in order to obtain a complete (up to date)
Standard Model expression for the decay rate, the following
known QCD contributions should also be included:
(i) The $O(\alpha_s^2)$ corrections due to the large mass splitting
within the t-b doublet \cite{Kni,Chettri};
(ii) The $O(\alpha_s^2)$ effects due to the virtual heavy quark
\cite{Sop,Knivert,Chetvirt};
(iii) The $O(\alpha_s^3)$ corrections coming from the triangle anomaly
type graphs in the limit $m_t \rightarrow \infty$ \cite{Lar}.
One also needs to include the electroweak corrections.
All those corrections can be found in \cite{Knirev}.

The calculated quark mass corrections to $O(\alpha_s^2)$ and
$O(\alpha_s^3)$ gave about 10-20\% corrections to the corresponding
massless results and are of marginal importance for the high precision
analysis at LEP. It is reasonable to expect that the missing
$O(\alpha_s^3)$ correction in the axial part will be of
the order similar to the corresponding vector part result.
However, the calculated mass corrections are important in
the low energy analysis, e.g., at PEP and PETRA
(or B-factory), where the vector part of eq.(17) is relevant.

For the Z$\rightarrow b\overline{b}$ decay mode, the $O(\alpha_s^2)$
mass corrections agree to the ones obtained in \cite{Chetm2V,Chetm2A}
using an indirect approach, based on the renormalization group arguments
and the results of \cite{Su,GKLS}. However, at the $O(\alpha_s^3)$, there
is a small disagreement. This, in fact, is due
to the incorrect numerical coefficient for $\zeta(3)$ term in \cite{Su},
which was used in \cite{Chetm2V}.
\footnote{There is also a misprint in eq.(23) for the
general expression for the $O(\alpha_s^3)$ term in \cite{Chetm2V}:
the division factor 92 should be replaced by 96.}
In the previous equations, the strong coupling $\alpha_s(M_Z)$
and the quark mass $m_f(M_Z)$ are understood as the $\overline{MS}$
quantities renormalized at the Z mass. The relation between the
$\overline{MS}$ running quark mass and the pole mass is derived
from the on shell results of \cite{BRH} (see \cite{H})
\begin{eqnarray}
\lefteqn{\hspace{-3cm} m_f^{(N)}(\mu)=m_f\biggl\{1
   -\frac{\alpha_s^{(N)}(\mu)}{\pi}
      \biggl(\frac{4}{3}+\log\frac{\mu^2}{m_f^2}\biggr)
      -\biggl(\frac{\alpha_s^{(N)}(\mu)}{\pi}\biggr)^2\biggl[K_f
       -\frac{16}{9}
        +\sum_{m_f<m_{f'}<\mu}\delta(m_f,m_{f'}) } \nonumber\\
 && \quad \hspace{5mm}
        +\biggl(\frac{157}{24}-\frac{13}{36}N
                           \biggr)\log\frac{\mu^2}{m_f^2}
      +\biggl(\frac{7}{8}-\frac{1}{12}N\biggr))\log^2\frac{\mu^2}{m_f^2}
                                                     \biggr]\biggr\},
\label{mMtopole}
\end{eqnarray}
where $m_f$ is the pole mass of the quark, the superscript $N$ indicates
that the corresponding quantity is evaluated for the $N$ active flavors
of quarks, $\mu$ is an arbitrary scale. (In the case of Z decay, $\mu=M_Z$
and $N=5$.)
\begin{equation}
K_f= 16.00650-1.04137N
  +\frac{4}{3}\sum_{m_l \leq m_f} \Delta\biggl(\frac{m_l}{m_f}\biggr),
 \hspace{3mm}
\delta(m_f,m_{f'})=-1.04137
    +\frac{4}{3}\Delta\biggl(\frac{m_{f'}}{m_f}\biggr)
\label{Kd}
\end{equation}
and the numerical values for the $\Delta$ at the relevant quark mass
ratios are given in \cite{H}. Numerically, in the case of
Z$\rightarrow b\overline{b}$ decay mode, $K_b\approx 12.5$
\footnote{slightly higher than the one given in \cite{Chetm2A}.}
and the sum over $m_{f'}$ drops out in eq.(\ref{mMtopole}).

The calculated mass corrections to the correlation functions are
relevant for the hadronic decay rates of the $\tau$ lepton.

\vspace{5mm}

\noindent
{\bf Acknowledgments}

It is a pleasure to thank D.~E.~Soper for
the collaboration, K.~G.~Chetyrkin, N.~G.~Deshpande, X.~G.~He, B.~A.~Kniehl,
M.~A.~Samuel and D.~M.~Strom for comments at various stages of the work.
This work was supported by the U.S. Department of Energy under
grant No. DE-FG06-85ER-40224.

\end{document}